\documentclass[pre, twocolumn, showpacs, superscriptaddress]{revtex4}
\usepackage{amsmath}
\usepackage{dcolumn}
\usepackage[xdvi]{graphicx}%
\makeatletter

\def\btt#1{\texttt{\@backslashchar#1}}%
\DeclareRobustCommand\bblash{\btt{\@backslashchar}}%
\makeatother

\newcommand{\bv}[1]{{\boldsymbol #1}}

\begin{document}

\title{Critical scaling of jammed system after quench of temperature}

\author{Michio Otsuki}
\affiliation{
Department of Physics and Mathematics, Aoyama Gakuin University,
5-10-1 Fuchinobe, Sagamihara, Kanagawa 229-8558, Japan}
\author{Hisao Hayakawa}
\affiliation{
Yukawa Institute for Theoretical Physics, Kyoto University,  Kitashirakawaoiwake-cho, Sakyo-ku, Kyoto 606-8502, Japan}

\begin{abstract}
Critical behavior of soft repulsive particles 
after quench of temperature near the jamming transition
is numerically investigated.
It is found that 
the plateau of the mean square displacement of tracer particles
and the pressure satisfy critical scaling laws. 
The critical density for the jamming transition
depends on the protocol to prepare the system,
 while the values of the critical exponents which are consistent with the prediction of a phenomenology 
are independent of the protocol.
\end{abstract}


\maketitle

\section{Introduction}


The jamming transition has attracted many physicists since Liu and Nagel indicated its similarity to the glass transition \cite{Liu}.
In naive sense, the glass transition is characterized by a divergence of time scale on a temperature-density plane \cite{Ediger,Angell,Debenedetti},
while the jamming transition is an athermal transition on a density-load plane
as the emergence of rigidity for
 materials such as granular materials, forms, and colloidal suspensions \cite{Jaeger,Durian,Pusey}.
 For frictionless particles,
it is known that the pressure, the elastic modulus,
and the characteristic frequency for the soft mode continuously emerge 
above a jamming transition point,
while the coordination number changes discontinuously at the point
\cite{OHern02, OHern03, Wyart,Majmudar}.

The jamming transition was discussed mainly on the axis of the density
without any load and temperature in some pioneer works
\cite{OHern02, OHern03, Wyart,Majmudar}.
It is instructive, however, that critical properties
have been clarified 
when we look at the behavior of the jamming on the density-load
plane in the zero load limit.
For instance, critical scaling laws for the rheological transition, similar
to those in continuous phase transitions, 
have been observed for sheared frictionless systems
\cite{Olsson,Hatano07,Hatano08,Tighe,Hatano10,Otsuki08,Otsuki09,Otsuki10,Tighe11,Otsuki12},
while the discontinuous transition and
the hysteresis loop are observed in the pressure and the shear stress
for sheared frictional granular materials \cite{Otsuki11,bob_nature}.

Coming back to the original idea in Ref.~\cite{Liu},
we can also discuss the jamming transition on the density-temperature plane, i.e. without any load.
This approach has an advantage to clarify the relationship between the glass transition~\cite{Ediger,Angell,Debenedetti} and the jamming transition~\cite{Jaeger,Durian,Pusey}, because the glass transition is originally defined only on the temperature-density plane.
So far the behavior on the density-temperature plane in the zero temperature limit
has been studied by some researchers, but the situation is still confusing.
Indeed, some indicated that
the critical fraction $\phi_{\rm G}$ for the divergence of the relaxation time
in the zero temperature limit
is identical to the critical point $\phi_{\rm J}$ for the jamming transition
of the athermal materials \cite{Cheng,Schweizer},
while the others suggested that two transition points are different 
\cite{Brambilla,Parisi,Krzakala,Berthier09,Berthier092}.
It is remarkable that Berthier and Witten have 
numerically confirmed from their simulation for soft repulsive particles at low temperature that 
(i) the relaxation time around the glass transition point $\phi_{\rm G}$
satisfies a scaling relation, and (ii) $\phi_{\rm G}$ 
is lower than that for the jamming point $\phi_{\rm J}$ \cite{Berthier09,Berthier092}.
It is also noticed that the separation between $\phi_{\rm G}$ and $\phi_{\rm J}$ 
is clearly demonstrated from a simulation for sheared soft spheres in the zero temperature and zero shear limits \cite{Ikeda}.

Recently, the critical behavior of repulsive particles 
on the density-temperature plane near the jamming transition
at zero temperature and high density
 has been studied both
numerically and theoretically
\cite{Jacquin, Berthier11,Brito09,Zhang}.
It is notable that the replica theory gives a prediction on both
the critical fraction and the critical exponents
 \cite{Jacquin, Berthier11}. 
The validity of their prediction for the critical behavior
of the pressure, the energy, and the divergence of the first peak 
of the radial distribution function
have numerically verified
 \cite{Jacquin, Berthier11}. 
However, it is unclear whether the critical exponents
are unique because 
they might
depend on the protocol to prepare the system
as for the critical density of the jamming transition
\cite{Chaudhuri,Vagberg}.

In this paper, 
to clarify critical behavior on the density-temperature plane
in the vicinity of the jamming transition point,
we numerically investigate
the value of plateau (VP) of the mean square displacement (MSD) 
of tracer particles and the pressure of
soft repulsive particles after quench of temperature
and demonstrate that the critical exponents for critical scaling laws
does not depend on the protocol, while the protocol dependence exists
in the critical fraction.
We should note that 
the pressure for soft spheres 
\cite{Berthier11}
and MSD for hard spheres \cite{Brito09}
have been numerically measured in the previous papers,
but this paper is the first report on the numerical study of MSD for soft spheres.

The organization of this paper is as follows.
In the next section, we will explain our set up and models.
In Sec. \ref{the mean square displacement:Sec},
we show the results of our simulation on the critical behavior for
VP and the pressure of the quenched soft particles.
In Sec. \ref{Critical_fraction},
we will show the jamming transition density depends
on the protocol to prepare the system.
In Sec. \ref{Critical:Sec}, 
we will present scaling laws for the plateau and the pressure,
and theoretically determine the critical exponents.
In Sec. \ref{Discussion:Sec}, we will discuss and conclude our results.

\section{Setup and Model}


We study a three dimensional system consists of $N$ soft spherical particles 
with  mass $m$ enclosed in a periodic cube of
linear size $L$. 
Note that the box size $L$ is fixed for the most cases, but is changed when we will determine the jamming point in Sec. IV.  
We prevent the system from crystallization by using a 50:50 binary mixture
of spheres of diameter ratio $1.4$
which is numerically confirmed from the radial distribution function,
where sharp peaks characterizing crystallization do not exist
\cite{OHern02,OHern03,Berthier09,Berthier092}.
It should be noted that the critical behavior for jamming transition of granular particles
is unchanged even for a mono-disperse system or a poly-disperse
system with equal number of particles of diameters $\sigma_0$, $0.9 \sigma_0$, $0.8 \sigma_0$, and $0.7 \sigma_0$ \cite{OHern02,OHern03,Otsuki09}.

For later convenience, let us use dimensionless quantities scaled by
$\sigma_0$ for the length, $m$ for the mass, and $\sqrt{m\sigma_0^2/\epsilon}$ for the time, respectively,
where we have introduced a characteristic energy scale $\epsilon$.
We assume that  
the interaction between $i$ and $j$ particles 
is described by a pair wise potential
\begin{equation}
V(r_{ij}) =  (1 - r_{ij} / \sigma_{ij})^2 
\theta \left (\sigma_{ij}  - r_{ij} \right ),
\label{Hook}
\end{equation}
where $\theta(x)$ 
 is the Heaviside step function satisfying
 $\theta(x) = 1$ for $x\ge0$ and
 $\theta(x) = 0$ for otherwise,
$r_{ij} = |\bv{r}_i - \bv{r}_j|$
and $\sigma_{ij} = (\sigma_i + \sigma_j)/2$
with the position $\bv{r}_i$ and the diameter $\sigma_i$ of the particle $i$.

We start from an equilibrium state at an initial temperature $T_{\rm I}$
and a volume fraction $\phi$. 
Then, we quench the system directly to a final temperature $T_{\rm F}$,
and the system subsequently evolves
at $T_{\rm F}$ by the velocity rescaling thermostat.
We use the system size $N=1000$.
We have checked the critical exponents do not change
when we use $N=4000$.
We adopt the leap-frog algorithm 
with the time interval $\Delta t = 0.01$.
We have verified that
the choice of the algorithm does not affect the average values of the pressure
and MSD within
the numerical accuracy when we use 
the velocity Verlet algorithm and $\Delta t = 0.001$.
We believe that the initial state is sufficiently equilibrated.
Indeed,  as long as we have checked, 
MSD exceeds $10$
and  we could not find any aging effects
during the equilibration process.
This system has been well studied in the previous papers
on the energy, the pressure, and the radial distribution function \cite{Jacquin,Berthier11}.

\section{Mean square displacement and pressure}
\label{the mean square displacement:Sec}

In this section, we summarize the results of our simulation on MSD and the pressure.
Note that the system has a fixed volumed fraction $\phi$ or a fixed volume in this section.

\begin{figure}
\begin{center}
\includegraphics[height=15em]{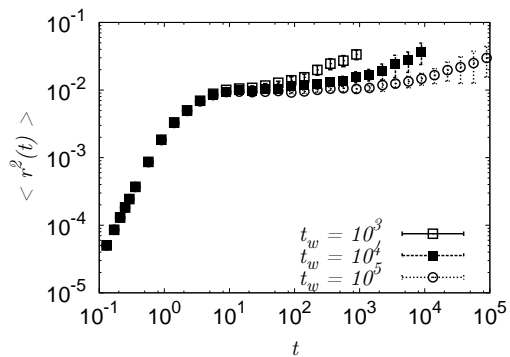}
\caption{
  MSD $\left <  r^2(t) \right >$ 
  as a function of time $ t$ for
$\phi=0.7$, $ T_{\rm I}=10^{-2}$ and $ T_{\rm F}=10^{-3}$
with
$ t_w = 10^3, 10^4$, and $10^5$.
}
\label{Msd_nu0.7_tw}
\end{center}
\end{figure}

First, let us consider the mean square displacement of larger 
tracer particles
at $T_{\rm F}$
\begin{equation}
\left < r^2 (t) \right > \equiv 
\sum_i^{N/2} \frac{\left < |\bv{r}_{\rm{L},i}(t + t_w) - \bv{r}_{\rm{L},i}(t_w) |^2 \right >}{N/2},
\end{equation}
where $\bv{r}_{\rm{L},i}(t + t_w)$ and $t_w$ are 
the position of the larger particle $i$ and the waiting time, i.e.,
the time elapsed after the quench, respectively.
Here, we ignore the displacement of the smaller particles
in order to eliminate the effect of rattlers \cite{OHern03}.
The bracket denotes an equilibrium ensemble average 
over the initial configurations.
In Fig. \ref{Msd_nu0.7_tw},
we plot MSD $\left <  r^2 (t) \right > $
as a function of the time $ t$ with
$\phi=0.7$, $ T_{\rm I}=10^{-2}$, $ T_{\rm F}=10^{-3}$
for the waiting time $ t_w = 10^3, 10^4$, and $10^5$,
where MSD exhibits clear plateaus.
The time to escape from the plateau
increases as the waiting time $t_w$ increases,
which indicates that the system does not 
reach an equilibrium state within the time window explored in our simulation
\cite{Kob}.
However, we should note that VP is independent of the waiting time $t_w$.

\begin{figure}
\begin{center}
\includegraphics[height=15em]{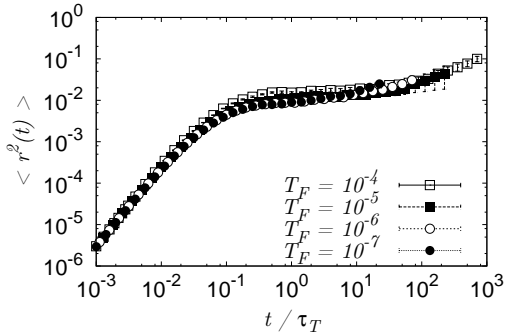}
\caption{
MSD $\left <  r^2(t) \right >$
  as a function of the time $ t$ 
  scaled by the thermal time $ \tau_T$ for
$\phi=0.62$, $ T_{\rm I}=10^{-2}$ and  $ t_w = 10^5$
with $ T_{\rm F} = 10^{-4}, 10^{-5}, 10^{-6}$, and $10^{-7}$.
}
\label{Msd_nu0.62}
\end{center}
\end{figure}

In Fig. \ref{Msd_nu0.62}, we plot MSD
 as a function of the time
$t$ divided by the ``thermal'' time 
$ \tau_T \equiv 1/\sqrt{ T_{\rm F}}$ for $\phi=0.62$, $ T_{\rm I}=10^{-2}$, $ t_w = 10^5$
with $ T_{\rm F}=10^{-4}, 10^{-5}, 10^{-6}$, and $10^{-7}$.
Thanks to the introduction of the scaled time $ t/  \tau_T$, 
MSD
for $\phi=0.62$
converges to a master curve,
which indicates that VP is almost independent of
the final temperature $T_{\rm F}$.
For relatively low density case,
it is known that the particles behave as a hard sphere liquid,
in which the dynamics is independent of the temperature
if the time is scaled by the thermal time
\cite{Berthier09,Berthier092}.
This is the reason for the scaling behavior as shown in Fig. \ref{Msd_nu0.62}.

\begin{figure}
\begin{center}
\includegraphics[height=15em]{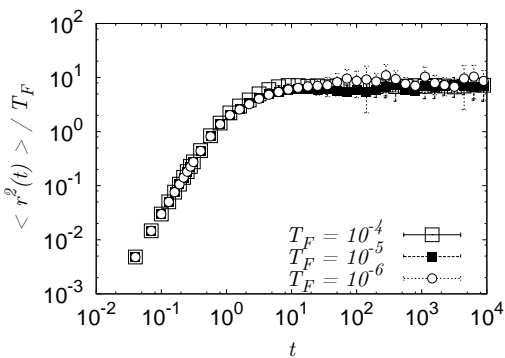}
\caption{
MSD $\left <  r^2(t) \right >$
scaled by $ T_{\rm F}$
  as a function of $ t$ for
$\phi=0.70$, $ T_{\rm I}=10^{-2}$ and $ t_w = 10^5$
with $ T_{\rm F}=10^{-4}, 10^{-5}$, and $10^{-6}$.
}
\label{Msd_nu0.70}
\end{center}
\end{figure}

On the contrary, MSD strongly depends on
$T_{\rm F}$ for denser cases.
In Fig. \ref{Msd_nu0.70},
we show $\left <  r^2(t) \right >$
scaled by $ T_{\rm F}$ as a function of the time $t$ for
$\phi=0.70$, $ T_{\rm I}=10^{-2}$, and $ t_w = 10^5$
with $ T_{\rm F}=10^{-4}, 10^{-5}$, and $10^{-6}$.
MSD $\left <  r^2(t) \right > $ scaled by $ 
T_{\rm F}$
converges to a  master curve,
which indicates that VP is proportional to $T_{\rm F}$.
For this case,
a particle is completely trapped within a cage and  fluctuates around
its equilibrium position.
The reason why VP is proportional to $T_F$ can be understood as follows.
The energy $\delta E$ due to the fluctuation of its position $\delta \bv{r}$
may be approximated as $\delta E \propto |\delta \bv{r}|^2$.
If we assume that the distribution of $\delta \bv{r}$ satisfies $\rho(\delta \bv{r}) \propto \exp( - \delta E / T_{\rm F})$,
$\rho(\delta \bv{r})$ depends on through $| \delta \bv{r} |^2 / T_{\rm F} $.
If we also assume that  VP is scaled by
the size of the fluctuation $\left < | \delta \bv{r} |^2 \right >$,
it is reasonable to obtain the scaling relation as in Fig. \ref{Msd_nu0.70}.

\begin{figure}
\begin{center}
\includegraphics[height=15em]{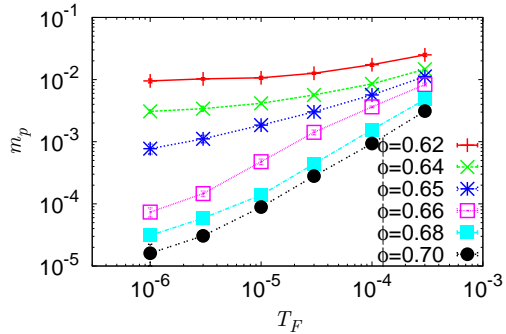}
\caption{
(Color online)
The value of plateau $ m_p$ 
  as a function of the $ T_{\rm F}$ for
$ T_{\rm I}=2.0 \times 10^{-3}$ and $ t_w = 10^5$
with $\phi = 0.62, 0.64, 0.65, 0.66, 0.68$, and $0.70$.
}
\label{mp}
\end{center}
\end{figure}

Here, let us introduce  $m_p$
as $\left < r^2(t) \right >$ at $ t= \tau_T$.
We should note that
$\langle r^2(t)\rangle$  changes less than $10$ \% 
for $ t>  \tau_T$.
Figure \ref{mp} exhibits
$ m_p$ 
as a function of
$ T_{\rm F}$ for $\phi = 0.62, 0.64, 0.65, 0.66, 0.68$ and $0.70$ with $ T_{\rm I} = 2.0 \times 10^{-3}$.
As we have noted, 
$m_p$ is a constant for lower densities
and is proportional to $T_{\rm F}$ for higher densities.
It is notable that $m_p$ behaves as a power-law function of $T_{\rm F}$ 
around $\phi=0.65$
\cite{OHern02}.

\begin{figure}
\begin{center}
\includegraphics[height=15em]{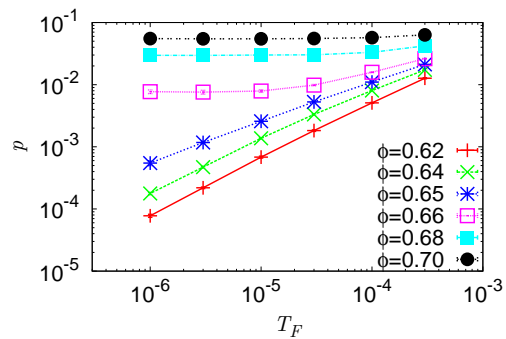}
\caption{
(Color online)
The pressure $ p$ 
  as a function of the final temperature $ T_{\rm F}$ 
  for $ T_{\rm I}=2.0 \times 10^{-3}$ and $ t_w = 10^5$
with $\phi = 0.62, 0.64, 0.65, 0.66, 0.68$, and $0.70$.
}
\label{p}
\end{center}
\end{figure}

The similar critical behavior can be observed for the pressure
at the final temperature $T_{\rm F}$
\begin{equation}
p = 
\frac{1}{3L^3} \left < \sum_i^N \sum_{j>i} r_{ij}
f(r_{ij})
\right >  
+
\frac{1}{3L^3} \left <   \sum_{i=1}^N \frac{ |\bv{p}_i|^2}{2m} \right >,
\end{equation}
where $\bv{p}_i$ is the momentum of the particle $i$
and $f(r_{ij}) \equiv - V_{ij}'(r_{ij})$ is the potential force.
Figure \ref{p} shows
the pressure $ p$
as a function of $ T_{\rm F}$
for $ T_{\rm I} = 2.0 \times 10^{-3}$.
For $\phi = 0.62$,
$p$ is almost proportional to $T_{\rm F}$
which is one of characteristic behavior of hard sphere liquids.
On the other hand,
$p$ is a constant at higher volume fraction such as $\phi = 0.70$,
because the pressure is determined by the rigidity of contact network of particles.
It is reasonable that the rigidity of the network is insensitive to the temperature near $T=0$.

\section{Protocol dependent critical fraction}
\label{Critical_fraction}

In this section, let us determine a critical fraction $\phi_{\rm J}$
after the quench. 
It should be noted that $\phi_{\rm J}$ is determined not by a simulation under a fixed volume but by a simulation by a compress or an expansion of the volume.

First, we prepare an equilibrium state of the volume fraction $\phi$
and the initial temperature $T_{\rm I}$.
Second, we quench the system directly to $T_{\rm F}$
and keep the temperature by the velocity rescaling thermostat.
Third, we further relax the system to the nearest potential energy minimum
by using the conjugate gradient technique \cite{LM}.
Then, if the pressure $p$ at the potential energy minimum
is higher than a threshold value $P_{\rm th}$,
we increase the volume per particle  $v$ by $\Delta v$.
Here, in order to increase the volume $v$, 
we change the system size $L$ and the position $\bv{r}_i$
of the $i$-th particle as $L \times \{(v + \Delta v)/v\}^{1/3}$
and $\bv{r}_i \times \{(v + \Delta v)/v\}^{1/3}$, respectively.
After the change of the volume, 
the system is relaxed to the nearest potential energy minimum.
We repeat the decrease of the volume fraction or 
expand the volume, and relax the system to a steady state.
Finally, the critical fraction $\phi_{\rm J}$ is determined 
from the volume per particle $v_{\rm J}$ 
where the the pressure becomes lower than $P_{\rm th}$
as $\phi_{\rm J} = v_{\rm av}/v_{\rm J}$ with $v_{\rm av}\equiv  \pi \sum_{i=1}^N \sigma_i^3/(6N)$.
If the pressure at the initial minimum of the potential energy 
is lower than $P_{\rm th}$,
we decrease the volume $v$ by $\Delta v$, 
relax the system, and repeat the decrease and the relaxation
until the pressure exceeds $P_{\rm th}$.
Then, we can determine the critical fraction $\phi_{\rm J}$
from the volume per particle $v_{\rm J}$ 
where the the pressure exceeds $P_{\rm th}$.
We use $ P_{\rm th} = 10^{-5}$ and $\Delta  v = 0.0005$.
We have checked that the critical fraction
does not change if we use
$ P_{\rm th} = 10^{-6}$ and $\Delta  v = 0.00005$.
It is also noted that the method to determine the critical fraction
is almost identical to that in the previous works 
\cite{OHern02, OHern03, Chaudhuri}.

\begin{figure}
\begin{center}
\includegraphics[height=15em]{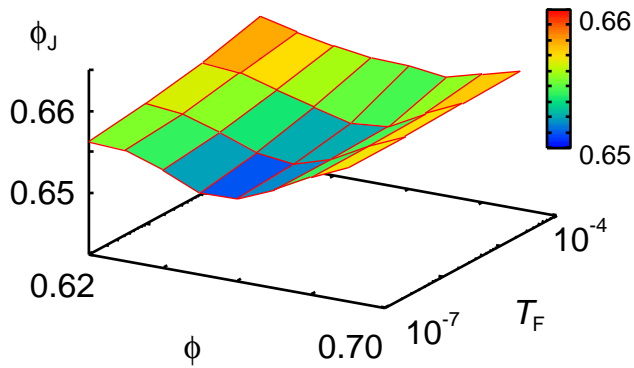}
\caption{(Color online)
The critical fraction 
$\phi_{\rm J}(\phi, T_{\rm F}, T_{\rm I})$
as a function of $ \phi$
and $ T_{\rm F}$
for the initial temperature $ T_{\rm I} = 2.0 \times 10^{-3}$.
}
\label{phij}
\end{center}
\end{figure}

\begin{figure}
\begin{center}
\includegraphics[height=15em]{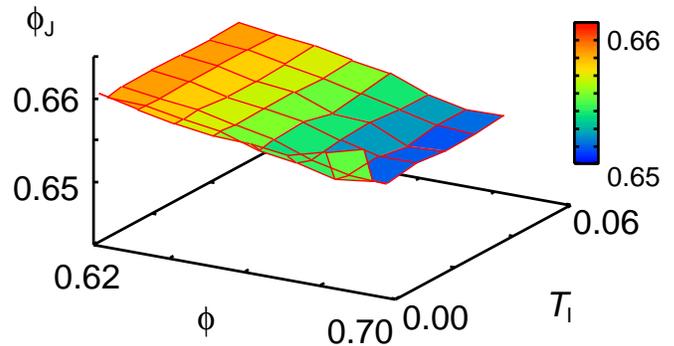}
\caption{(Color online)
The critical fraction 
$\phi_{\rm J}(\phi, T_{\rm F}, T_{\rm I})$
as a function of 
$ T_{\rm I}$
and 
$ \phi$ 
for $ T_{\rm F} = 2.0 \times 10^{-3}$.
}
\label{phij_Ti}
\end{center}
\end{figure}

In Fig. \ref{phij}, we display the critical fraction 
$\phi_{\rm J}(\phi, T_{\rm F}, T_{\rm I})$
as a function of $ \phi $ 
and $ T_{\rm F}$
for $ T_{\rm I}=2.0 \times 10^{-3}$.
Figure \ref{phij_Ti} exhibits
$\phi_{\rm J}(\phi, T_{\rm F}, T_{\rm I})$
as a function of the initial temperature $ T_{\rm I}$
and $ \phi $ 
for $ T_{\rm F} = 1.0 \times 10^{-4}$.
These figures reveal that the critical fraction $\phi_{\rm J}$
depends on the initial equilibrium state, i. e. $T_{\rm I}$ and $\phi$, and the 
quenched state at $T_{\rm F}$. 
We note that the existence of the initial state dependence
has already numerically demonstrated in  Ref. \cite{Chaudhuri},
but the dependence on the quenched state at $T_{\rm F}$ within our knowledge has not been discussed in any other papers.

\section{Critical scalings of the value of plateau and the pressure}
\label{Critical:Sec}

In this section, let us develop the scaling analysis to characterize the behavior of MSD and the pressure.
This section consists of three parts. 
In the first part, we summarize some asymptotic relations in the scaling functions.
In the second part, we briefly introduce the method to evaluate the scaling exponents.
In the last part, we discuss the values of the critical exponents.

Through our simulation, we have confirmed that
that the value of plateau $m_p(\phi,T_{\rm F}, T_{\rm I})$
and the pressure $p( \phi,T_{\rm F},T_{\rm I})$ satisfy the scaling laws
with the protocol dependent critical fraction $\phi_{\rm J}(\phi,T_{\rm F}, \phi)$:
\begin{eqnarray}
m_p(\phi,T_{\rm F}, T_{\rm I}) & = & T_{\rm F} ^{a_m} {\mathsf M} \left (\frac{\phi - \phi_{\rm J}(\phi,T_{\rm F}, T_{\rm I})}{T_{\rm F}^b}
\right ), \label{mp:eq1}\\
p(\phi,T_{\rm F}, T_{\rm I}) & = & T_{\rm F}^{a_p} {\mathsf P} \left (\frac{\phi - \phi_{\rm J}(\phi,T_{\rm F}, T_{\rm I})}{T_{\rm F}^b}
\right ), \label{p:eq1}
\end{eqnarray}
where
$a_m$, $a_p$, and $b$ are the critical exponents.
Here, we assume that the scaling functions ${\mathsf M} (x)$ and ${\mathsf P} (x)$ satisfy
\begin{eqnarray}
\lim_{x \to \infty} {\mathsf M} (x) & \propto & x^{(a_m-1)/b}, \label{sfunc1} \\
\lim_{x \to -\infty} {\mathsf M} (x) & \propto & |x|^{a_m/b}, \label{sfunc2} \\
\lim_{x \to \infty} {\mathsf P} (x) & \propto & x^{a_p/b}, \label{sfunc3} \\
\lim_{x \to -\infty} {\mathsf P} (x) & \propto & |x|^{(a_p-1)/b}, \label{sfunc4}
\end{eqnarray}
because of the relations
\begin{eqnarray}
\lim_{T_{\rm F} \to 0} m_p  & = & F_1(\phi - \phi_{\rm J}), \\
\lim_{T_{\rm F} \to 0} p  & = & T_{\rm F} F_2(\phi - \phi_{\rm J}),
\end{eqnarray}
for $\phi < \phi_{\rm J}$, and
\begin{eqnarray}
\lim_{T_{\rm F} \to 0} m_p  & = & T_{\rm F}F_3(\phi - \phi_{\rm J}), \\
\lim_{T_{\rm F} \to 0} p  & = & F_4(\phi - \phi_{\rm J}),
\end{eqnarray}
for $\phi > \phi_{\rm J}$,
where $F_1$, $F_2$, $F_3$, and $F_4$ are functions depending only on
$\phi - \phi_{\rm J}$.
The corresponding scaling forms have already discussed in terms of the replica theory
\cite{Jacquin, Berthier11}.
The similar critical scaling laws are also found for the jamming transition
for sheared frictionless particles 
\cite{Olsson,Hatano07,Hatano08,Tighe,Hatano10,Otsuki08,Otsuki09,Otsuki10}.

\begin{figure}
\begin{center}
\includegraphics[height=15em]{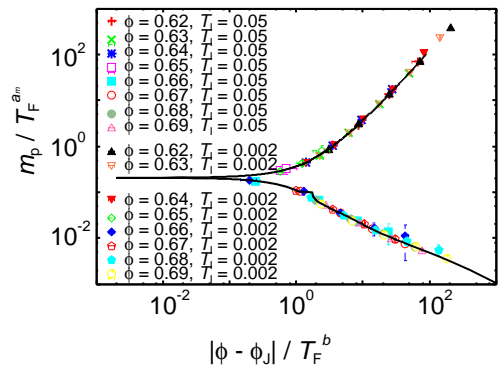}
\caption{
(Color online)
  Scaling plots of 
$ m_p$ characterized by Eq. \eqref{mp:eq1}
for $ T_{\rm I}=2.0 \times 10^{-3}$ and $5.0 \times 10^{-2}$ with 
  $ t_w = 10^5$ and
$\phi = 0.62, 0.63, 0.64, 0.65, 0.66, 0.67, 0.68$, and $0.69$.
The solid line is the scaling function given by Eq. \eqref{sfunc:M}
with the exponents estimated as Eq. \eqref{exponent:numerical}.
The values of the other fitting parameters are shown in the text.
}
\label{mp_scale}
\end{center}
\end{figure}

\begin{figure}
\begin{center}
\includegraphics[height=15em]{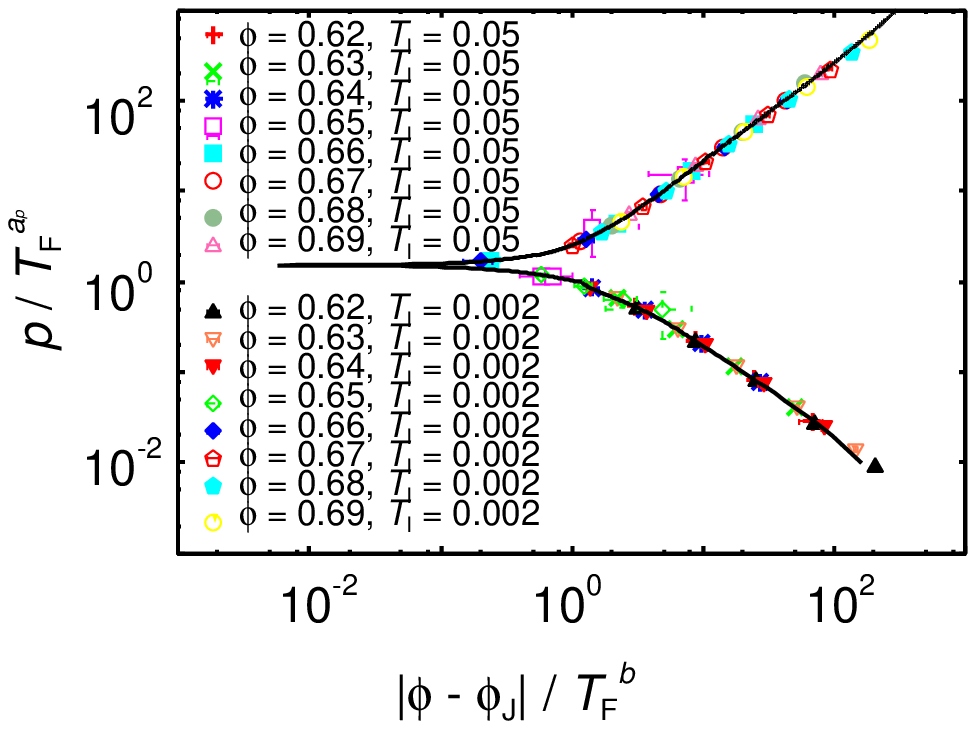}
\caption{
(Color online)
  Scaling plots of $ p$ characterized by Eq. \eqref{p:eq1}
for $ T_{\rm I}=2.0 \times 10^{-3}$ and $5.0 \times 10^{-2}$ with $ t_w = 10^5$ and
$\phi = 0.62, 0.63, 0.64, 0.65, 0.66, 0.67, 0.68$, and $0.69$.
The solid lines are the scaling function given by Eq. \eqref{sfunc:M}
with the exponents Eq. \eqref{exponent:numerical}.
The values of the other fitting parameters are shown in the text.
}
\label{p_scale}
\end{center}
\end{figure}

Figures \ref{mp_scale} and \ref{p_scale}
show the scaling plots based on Eqs. \eqref{mp:eq1} and \eqref{p:eq1}
for $ T_{\rm I} = 2.0 \times 10^{-3}$ and $5.0 \times 10^{-2}$,
respectively.
These figures confirm the validity of Eqs. \eqref{mp:eq1} and \eqref{p:eq1}.
Here, we numerically estimate 
\begin{equation}
a_m              = 0.722 \pm 0.004, \qquad a_p              = 0.506 \pm 0.004, \qquad
b               = 0.471 \pm 0.004
\label{exponent:numerical}
\end{equation}
for different initial temperatures $ T_{\rm I} = 2.0 \times 10^{-3},  5.0 \times 10^{-3},  1.0 \times 10^{-2},  2.0 \times 10^{-2},  3.0 \times 10^{-2},  4.0 \times 10^{-2}$, and $4.0 \times 10^{-2}$
by using the Levenberg-Marquardt algorithm \cite{LM},
where we expand the functional forms of the scaling functions as
\begin{eqnarray}
{\mathsf M}^{-1} (x) & = & 
\left\{
  \begin{array}{ll}
  \sum_{n=0}^5 A_n  \log(x) ^n & (x \ge 1), \\
  \sum_{n=0}^5 B_n  \log(x) ^n & (x < 1),
\end{array} 
  \right.
\label{sfunc:M}\\
{\mathsf P}^{-1} (x) & = & 
\left\{
  \begin{array}{ll}
  \sum_{n=0}^5 C_n  \log(x) ^n & (x \ge 1), \\
  \sum_{n=0}^5 D_n  \log(x) ^n & (x < 1),
\end{array} 
  \right.
\label{sfunc:P}
\end{eqnarray}
with fitting parameters $A_n$, $B_n$, $C_n$, and $D_n$.
Here, we estimate the values of the fitting parameters as $(A_0, A_1, A_2, A_3, A_4, A_5) = (1.3, -5.2, 4.0, -6.2, 2.6, -0.8)$,
$(B_0, B_1, B_2, B_3, B_4, B_5) = (2.2, 5.2, 16, -74, -167, -113)$,
$(C_0, C_1, C_2, C_3, C_4, C_5) = (-0.9, -8.0, -12, 19, -11, 4.1)$,
$(D_0, D_1, D_2, D_3, D_4, D_5) = (-1.2, 0.4, -32, -42, -36, -5.8)$.
This method has been used to estimate
the critical exponents for the jamming transition for sheared frictionless particles
\cite{Olsson11} and for sheared frictional grains \cite{Otsuki11}.
As shown in Figs. \ref{mp_scale} and \ref{p_scale},
$m_p$ and $p$ for different initial temperature $T_{\rm I}$
satisfy the critical scalings with the same critical exponents.
This indicates that the critical exponents
are independent of the protocol although the critical fraction
depends on it as demonstrated in the previous section.

Now, let us estimate the critical exponents in Eqs. \eqref{mp:eq1}
and \eqref{p:eq1}
by using the previous phenomenological results on the jammed soft particles
without temperature \cite{OHern02,OHern03}, and the unjammed hard spheres
\cite{Salsburg, Brito09}.
From Eqs. \eqref{mp:eq1} -\eqref{sfunc4},
we readily obtain
\begin{eqnarray}
\lim_{T_{\rm F} \to 0} m_p  & \propto & |\phi-\phi_{\rm J}|^{a_m/b}, 
\label{unjam_mp}\\
\lim_{T_{\rm F} \to 0} p  & \propto & T_{\rm F}|\phi-\phi_{\rm J}|^{(a_p-1)/b},
\label{free_volume}
\end{eqnarray}
for $\phi < \phi_{\rm J}$, and
\begin{eqnarray}
\lim_{T_{\rm F} \to 0} m_p  & \propto &  T_{\rm F}|\phi-\phi_{\rm J}|^{(a_m-1)/b}, 
\label{jam_mp}\\
\lim_{T_{\rm F} \to 0} p  & \propto &  |\phi-\phi_{\rm J}|^{a_p/b},
\label{jam}
\end{eqnarray}
for $\phi > \phi_{\rm J}$.

For $\phi<\phi_{\rm J}$, 
the pressure may satisfy
\begin{eqnarray}
p  & \propto & T_{\rm F}|\phi-\phi_{\rm J}|^{-1}
\label{p:scale1}
\end{eqnarray}
as suggested by the free volume theory for hard sphere liquids \cite{Salsburg}.
From the comparison of this equation with Eq. \eqref{free_volume},
we obtain
\begin{eqnarray}
\frac{a_p-1}{b} = -1.
\label{exponent:1}
\end{eqnarray}

For $\phi>\phi_{\rm J}$,
the pressure might be given by
\cite{OHern02, OHern03}
\begin{eqnarray}
p  & \propto &|\phi-\phi_{\rm J}|.
\label{p:scale2}
\end{eqnarray}
From Eqs. \eqref{jam} and \eqref{p:scale2},
we obtain
\begin{eqnarray}
\frac{a_p}{b} = 1.
\label{exponent:2}
\end{eqnarray}

For $\phi<\phi_{\rm J}$,
MSD for hard sphere liquids is expected to satisfy
\cite{Brito09}
\begin{eqnarray}
m_p & \sim &
|\phi - \phi_{\rm J}|^{3/2}.
\label{unjam:eq}
\end{eqnarray}
Thus, from Eqs. \eqref{unjam_mp} and \eqref{unjam:eq},
we obtain
\begin{equation}
\frac{a_m}{b} = \frac{3}{2}.
\label{exponent:3}
\end{equation}

From Eqs. \eqref{exponent:1}, \eqref{exponent:2}, and \eqref{exponent:3},
we obtain the critical exponents
\begin{equation}
a_m = \frac{3}{4}, \qquad a_p = b = \frac{1}{2},
\label{exponent:theory}
\end{equation}
which are not far from the numerical estimated exponents 
presented in Eq. \eqref{exponent:numerical}.

\section{Discussion and conclusion}
\label{Discussion:Sec}


Now, let us discuss and conclude our results.
First, we discuss the relationship between our approach and the papers by Berthier and Witten \cite{Berthier09,Berthier092}.
Second, we compare our results with the prediction by the replica theory \cite{Berthier11}.
Third, we comment on the possibility to extend our model to another model of contact.
In final, we summarize our results.


The previous papers \cite{Berthier09,Berthier092} demonstrate that
the structural relaxation time satisfies a scaling relation
around $\phi_{\rm G} = 0.635$.
We could also reproduce their scaling in our simulation,
though the results are not reported in this paper.
The scaling relation means that
the time to escape from the plateau of $\left < r^2(t) \right >$
for hard sphere liquids diverges at $\phi_{\rm G}$,
but the value of plateau does not exhibit any criticality around $\phi_{\rm G}$
because the particles can move in the cage
even at $\phi_{\rm G}$.
Moreover, the pressure continuously changes around 
$\phi_{\rm G}$ because the divergence of the relaxation time 
is not related to the pressure.
Hence, $\phi_{\rm G}$ does not appear in the scaling relations 
\eqref{mp:eq1} and \eqref{p:eq1}.
From Eq. \eqref{mp:eq1}, $m_p$ 
for hard sphere liquids becomes zero at $\phi_{\rm J}$,
which indicates that the dynamics of the particles in the cage 
is frozen at $\phi_{\rm J}$ \cite{Brito09}.


In Ref. \cite{Berthier11},
the replica analysis is used for the explanation of the jamming transition on the temperature-density plane for harmonic spheres.
They derived 
the identical scaling relation \eqref{p:eq1} for the pressure
with the critical exponents $a_p = b = 1/2$ corresponding to Eq. \eqref{exponent:theory}
and
our numerical results in Eq. \eqref{exponent:numerical}.
In addition, they suggested
a critical relation
\begin{eqnarray}
A^* & = & T ^{\gamma} {\mathsf A}' \left (\frac{\phi - \phi_{\rm J}}{T^\nu}
\right )
\end{eqnarray}
for the optimal cage size $A^*$
with critical exponents $\gamma = \nu = 1/2$.
Because both of $A^*$ and $m_p$ are the characteristic length,
it may be reasonable that they satisfy the identical critical scaling
if we assume that there exists only one characteristic length scale,
but the values of the critical exponents $\gamma = \nu = 1/2$ from the replica theory differ from
those for  $m_p$ in Eq. \eqref{exponent:numerical}.

In this paper, we only consider the system with
Hookean soft-core repulsion given by Eq. \eqref{Hook}.
For a system with the interaction potential
\begin{equation}
V(r_{ij}) =  (1 - r_{ij} / \sigma_{ij})^ {\Delta + 1}
\theta \left (\sigma_{ij}  - r_{ij} \right )
\label{potential'}
\end{equation}
with a exponent $\Delta$, the scaling of the pressure
given by Eq. \eqref{p:scale2} is expected to be changed as
\cite{OHern02, OHern03}
\begin{eqnarray}
p  & \propto &|\phi-\phi_{\rm J}|^\Delta,
\end{eqnarray}
which leads to 
\begin{eqnarray}
\frac{a_p}{b} = \Delta.
\label{exponent:2'}
\end{eqnarray}
From Eqs. \eqref{exponent:1}, \eqref{exponent:3}, and \eqref{exponent:2'},
the critical exponents for the system with the potential
given by Eq. \eqref{potential'} are expected to be
\begin{equation}
a_m = \frac{3}{2\Delta + 2}, \qquad a_p = \frac{\Delta}{\Delta + 1}, \qquad  b = \frac{1}{\Delta + 1}.
\end{equation}
The similar dependence of the critical exponents is confirmed 
in the sheared granular systems \cite{Otsuki09}.

In conclusion, we have numerically investigated critical behavior of VP of
MSD and the pressure
for soft repulsive particles after quench
near the jamming transition point.
We verify the existence of the critical scaling relations \eqref{mp:eq1} and \eqref{p:eq1},
and numerically evaluate the critical exponents and the critical fraction.
The critical fraction exhibits the protocol dependence,
while the critical exponents are independent of the protocol,
which are close to the estimation Eq. \eqref{exponent:theory} in terms of the combination of the existing arguments.


\begin{acknowledgments}
We thank G. Szamel, S. Teitel, K. Miyazaki and L. Berthier
for valuable discussions.
This work is partially supported by the 
Ministry of Education, Culture, Science and Technology (MEXT), Japan
 (Grant Nos. 21540384 and 22740260) and the Grant-in-Aid for the global COE program
"The Next Generation of Physics, Spun from Universality and Emergence"
from MEXT, Japan.
The numerical calculations were carried out on Altix3700 BX2 at 
the Yukawa Institute for Theoretical Physics (YITP), Kyoto University.
\end{acknowledgments}

\end{document}